\documentclass[final,5p,times,twocolumn]{elsarticle}

\usepackage{graphicx}
\usepackage{amssymb}

\journal{Physica E}

\begin{document}

\begin{frontmatter}
\date{\today}

 \author[1,2]{P. P. Baruselli}
 \ead{barusell@sissa.it}
 \author[2,3,4,5]{A. Smogunov}
 \author[1,3,6]{M. Fabrizio}
 \author[1,3,6]{E. Tosatti}

\address[1]{SISSA, Via Beirut 2/4, Trieste 34014, Italy}
\address[2]{CNR-IOM Democritos, Via Beirut 2/4, Trieste 34014, Italy}
\address[3]{ICTP, Strada Costiera 11, Trieste 34014, Italy}
\address[4]{Voronezh State University, University Square 1, Voronezh 394006, Russia}
\address[5]{present address: CEA Saclay, France}
\address[6]{INFM, Democritos Unit\'a di Trieste, Via Beirut 2/4, Trieste 34014, Italy}

\title{Kondo effect of magnetic impurities on nanotubes}

\begin{abstract}
The effect of magnetic impurities on the ballistic conductance of nanocontacts is, as
suggested in recent work, amenable to ab initio study \cite{naturemat}. Our method proceeds via
a conventional density functional calculation of spin and symmetry dependent
electron scattering phase shifts, followed by the subsequent numerical
renormalization group solution of Anderson models -- whose ingredients and
parameters are chosen so as to reproduce these phase shifts. We apply this method to
investigate the Kondo zero bias anomalies that would be caused in the ballistic
conductance of perfect metallic (4,4) and (8,8) single wall carbon
nanotubes, ideally connected to leads at the two ends, by externally
adsorbed Co and Fe adatoms. The different spin and electronic structure
of these impurities are predicted to lead to a variety of Kondo
temperatures, generally well below 10 K, and to interference between channels leading to
Fano-like conductance minima at zero bias.
\end{abstract}


\end{frontmatter}

\section{Introduction}\label{sec1}

The zero-bias anomalies observed in STS conductance spectroscopy 
through adsorbed magnetic impurities and to some extent in metal break
junctions have recently revived interest in the Kondo effect. Addressing
these systems theoretically poses several problems. In the first
place, and unlike quantum dots, ab initio electronic structure calculations 
such as density functional theory (DFT) are essential to establish a 
quantitatively meaningful starting point. Which among the impurity-related 
levels and resonances drive the spin polarization, what is their multiplicity,
their hybridization, etc. are all questions that need an ab initio calculation. 
Next, this information must be translated into some manageable many body 
Hamiltonian, possibly without the loss of the brute quantitative information 
provided by DFT. Finally, the many body Hamiltonian(s) must be solved, 
to extract Kondo parameters and the predicted conductance features near 
zero bias, possibly with their behavior with parameters such as nanocontact 
geometry, temperature and external field, to be eventually compared with
experiment. One approach in this direction was recently taken by our
group \cite{naturemat}. Given a nanocontact between two leads, one identifies, with the 
help of symmetry, the impinging and outgoing channels that carry
current across the impurity. From the matching symmetry selected local
densities of states at the impurity, one identifies the important
impurity orbitals with their different magnetic splittings and
hybridizations. This leads to formulate multi-orbital Anderson models,
which contain a multiplicity of parameters to be adjusted. In our scheme 
the parameters are adjusted to yield, within the Hartree-Fock 
approximation, the same channel- and spin-dependent impurity scattering phase 
shifts as those that we calculate ab initio by DFT -- whose input 
information is therefore put to maximal use. For the last step, solving the Anderson
models, we employed a standard numerical renormalization group (NRG) scheme.
While other groups have dealt with the overall Kondo problem in different
ways \cite{jacob2009, costi2009}, we find our ``DFT + NRG'' route extremely 
instructive, and worth exploring in more complex situations than the simple
Au-Ni-Au contact studied in Ref \cite{naturemat}. 
In the present application we  consider a single wall carbon nanotube (SWNT)
as our linear conducting system, and a single externally adsorbed transition metal atom,  
either Co or Fe, as the  magnetic impurity. To begin with, the metallic nanotube has two 
conducting channels instead of only one as Au. The magnetic atoms
in turn have in principle a richer multiplicity of magnetic levels
than Ni. We wish to explore what this richness might bring.

\section{Systems and symmetries}\label{sec2}
\begin{figure}\centering
\includegraphics[scale=0.3]{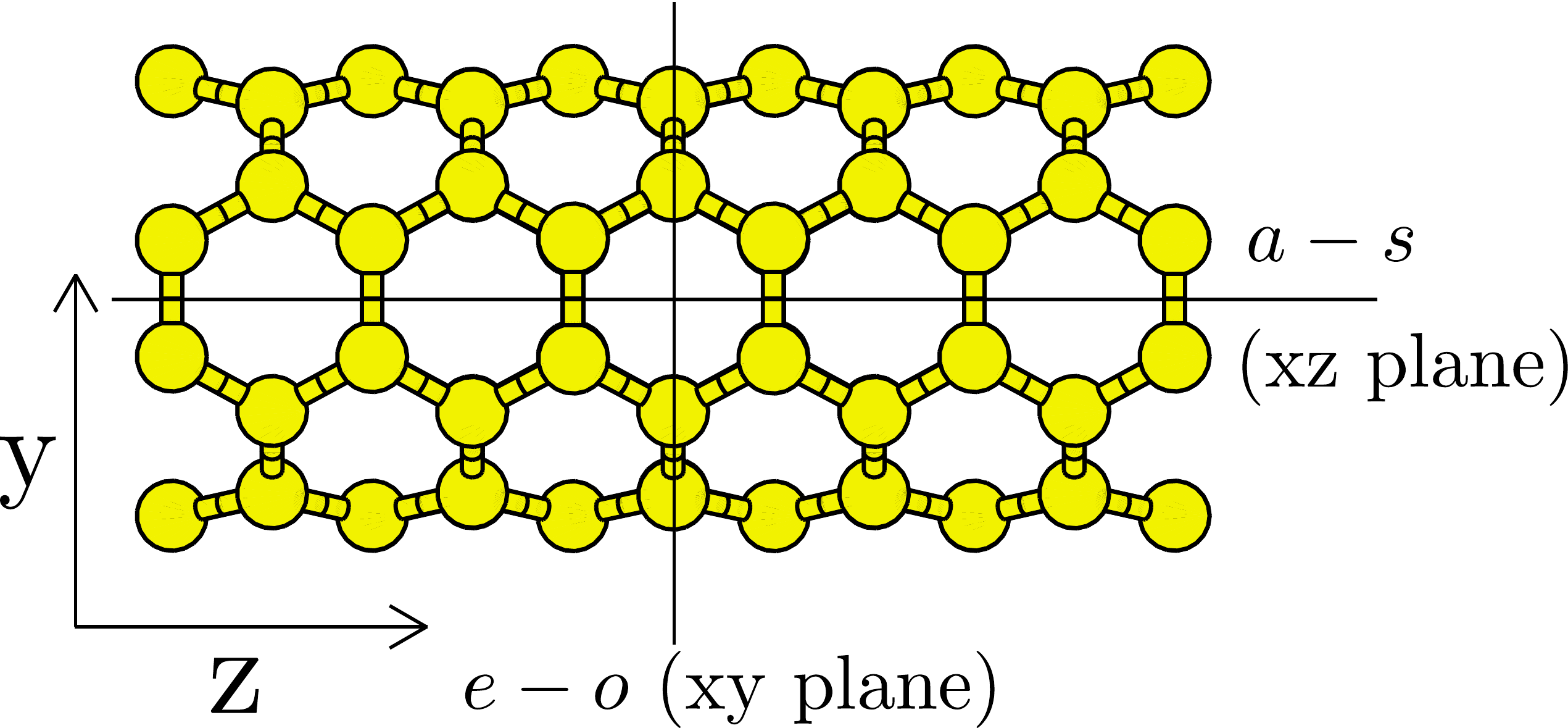}
\includegraphics[scale=0.3]{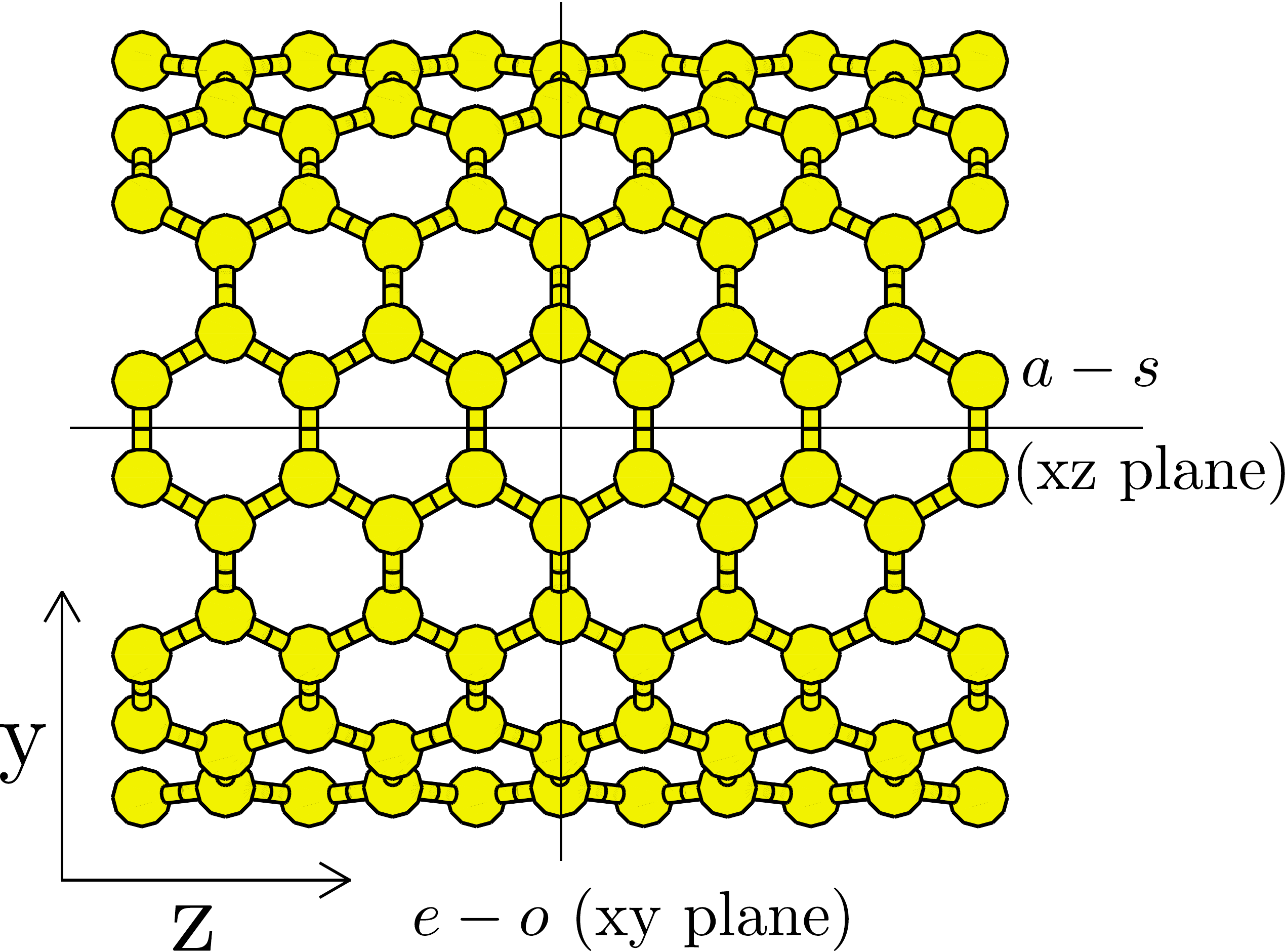}
\caption{A schematic view of clean (4,4) and (8,8) SWNTs with their symmetries.}\label{fig1} 
\end{figure}

\begin{figure}\centering
\includegraphics[scale=0.3]{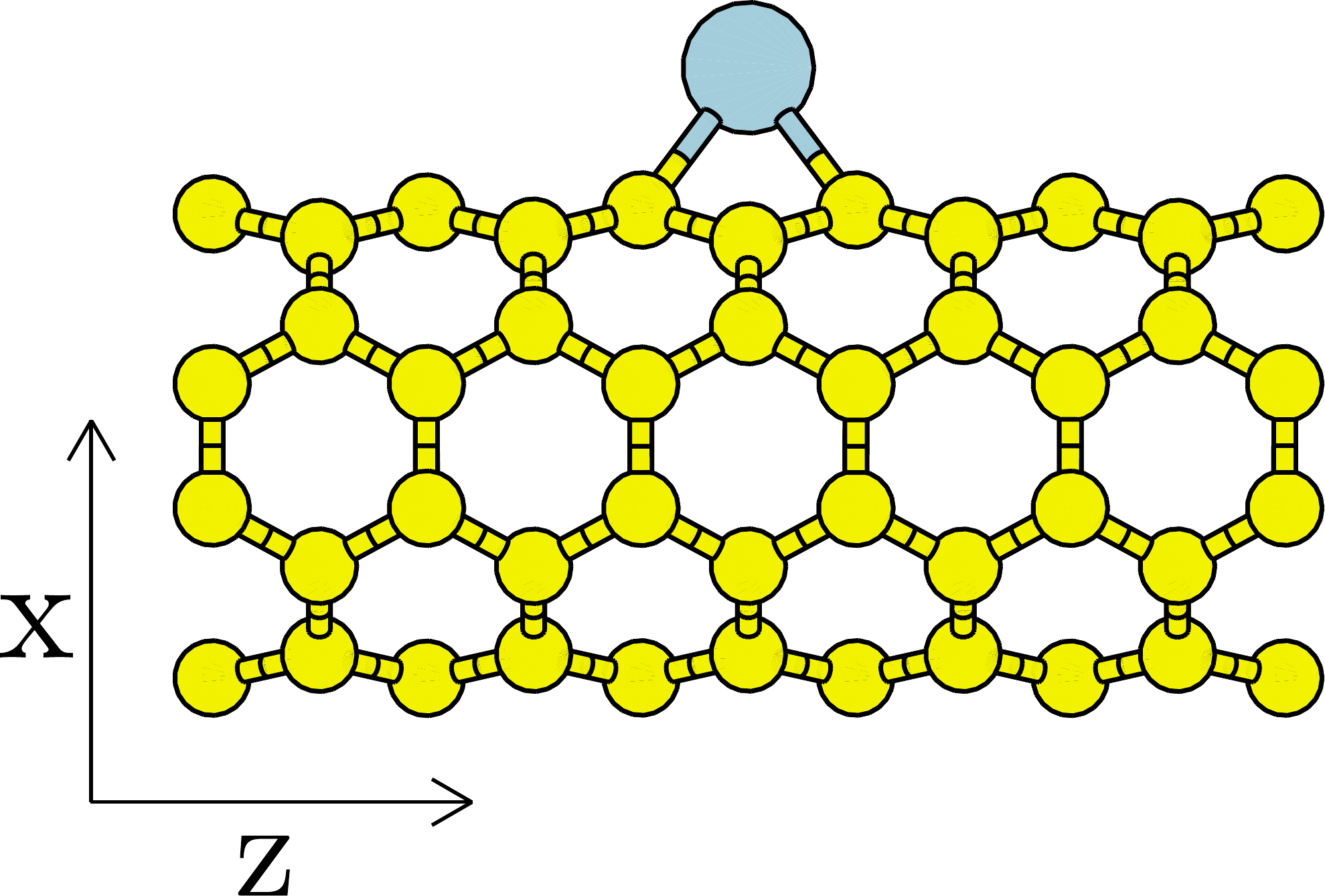}
\includegraphics[scale=0.3]{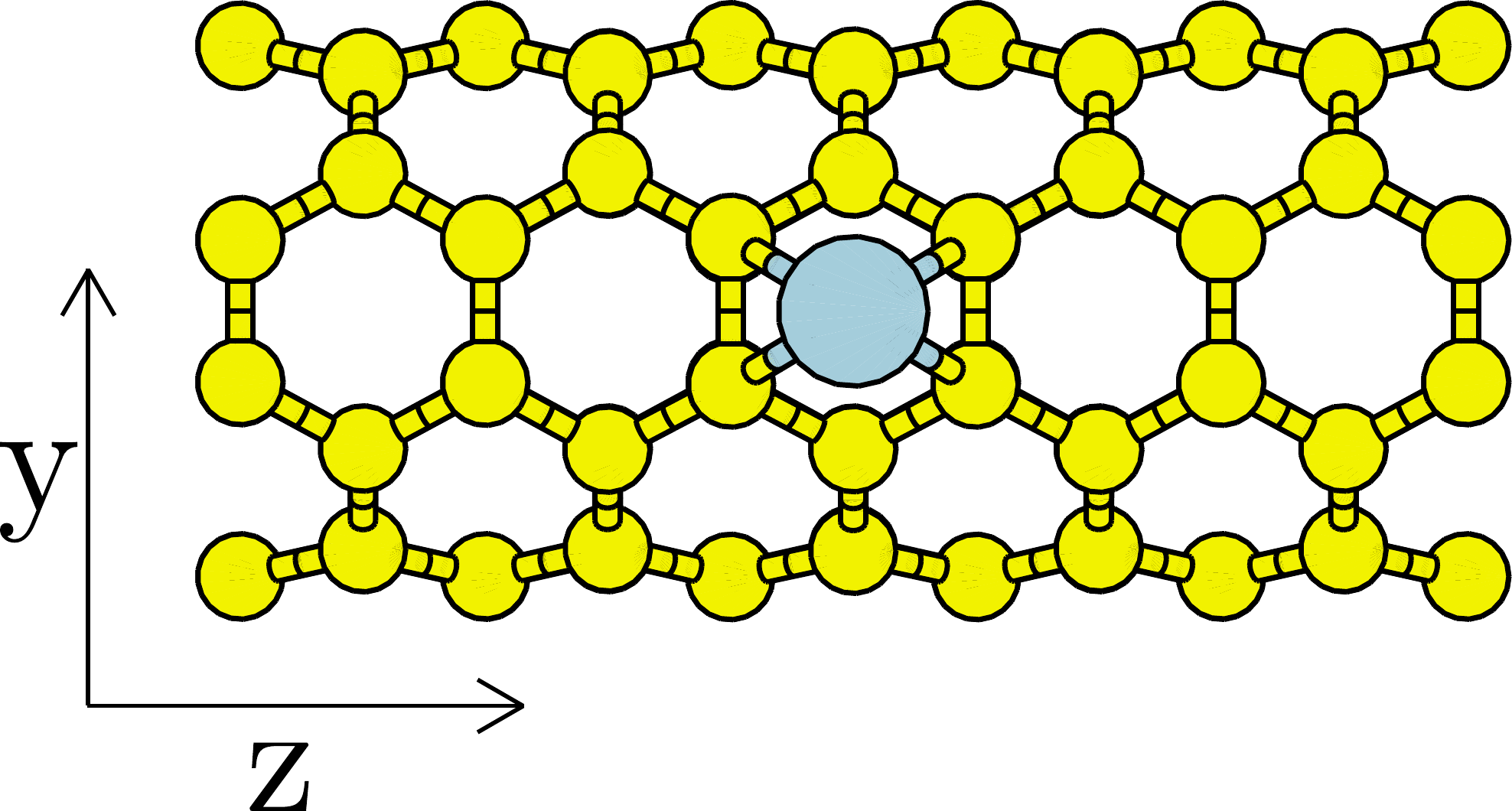}
\includegraphics[scale=0.3]{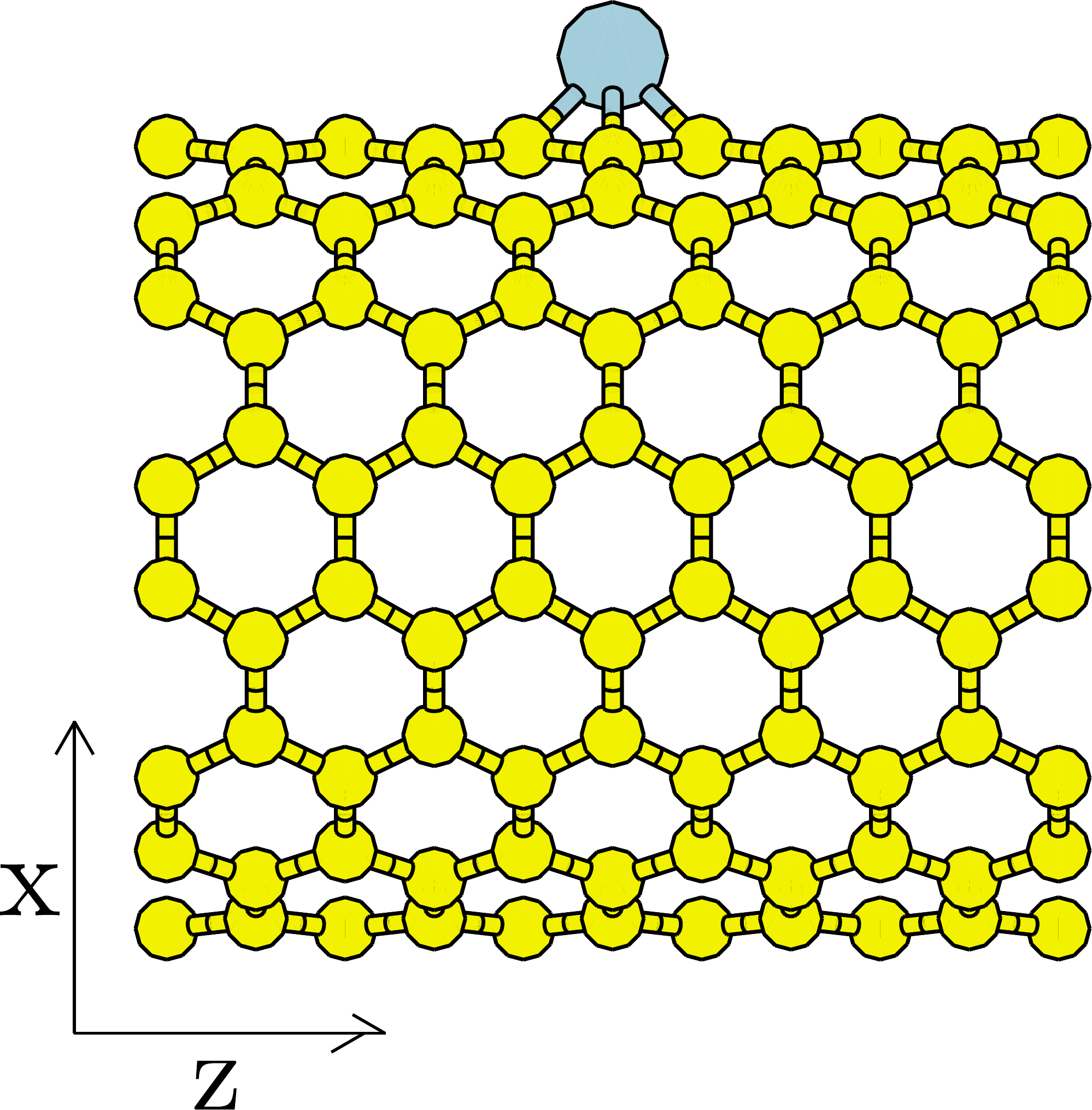}
\includegraphics[scale=0.3]{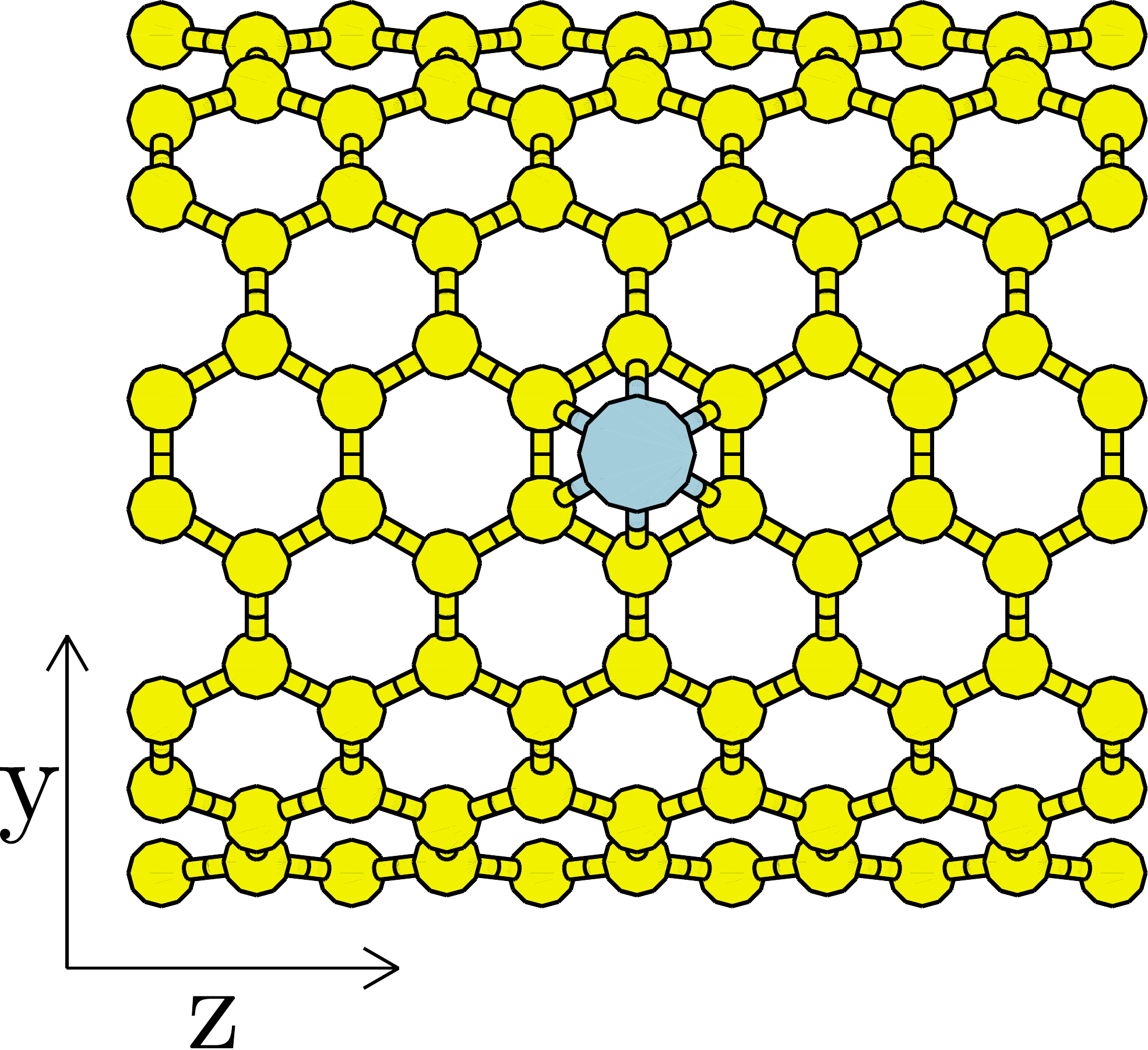}
\caption{A schematic view of (4,4) and (8,8) SWNTs with an impurity adsorbed in the \textit{hollow} position.} \label{fig2}
\end{figure}

We considered alternatively Co or Fe impurities on either (4,4) or (8,8) metallic
SWNTs (see fig. \ref{fig1}). If $z$ is the SWNT axis, its electronic states 
of can be classified according to parity with respect to  $xy$ plane reflection 
($e-o$, even- odd)  and $xz$ plane reflection ($s-a$, symmetric and antisymmetric).
DFT calculations (see section \ref{sec3}) predict that the externally 
adsorbed impurities should have minimum energy when at the \textit{hollow}
site (see fig. \ref{fig2}), that is above the center of a
carbon hexagon. Assuming that geometry, the impurity electronic states can be 
classified according to the same parity numbers as those of clean
SWNTs. We are interested in particular in $3d$ and $4s$ impurity
orbitals, whose parities are shown in tab. \ref{tab1}.

\section{Ab initio electronic structure}\label{sec3}
We carried out standard density-functional theory (DFT) calculations,
allowing for full relaxation of all atomic positions in a unit cell,
which comprised 80 and 160 
carbon atoms for the (4,4) and (8,8) tubes
respectively plus one Co or Fe adsorbed impurity. Calculations used 
the standard plane-wave package Quantum-ESPRESSO \cite{qe} within the generalized gradient
approximation (GGA) to exchange-correlation functionals in parametrization of
Perdew, Burke and Ernzerhof. The plane wave cut-offs were 
30 Ry and 300 Ry for the wave functions and for the charge density, 
respectively. Integration over the one-dimensional Brillouin zone was accomplished using 
$8$ k-points and a smearing parameter of $10$ mRy. When necessary to test the
sensitivity of DFT results to correlation effects, we extended to ``GGA+U''  with
a reasonably small Hubbard ``U'' \cite{qe}-- but generally the straight DFT
result was used.

\begin{figure}\centering
\includegraphics[width=0.45\textwidth]{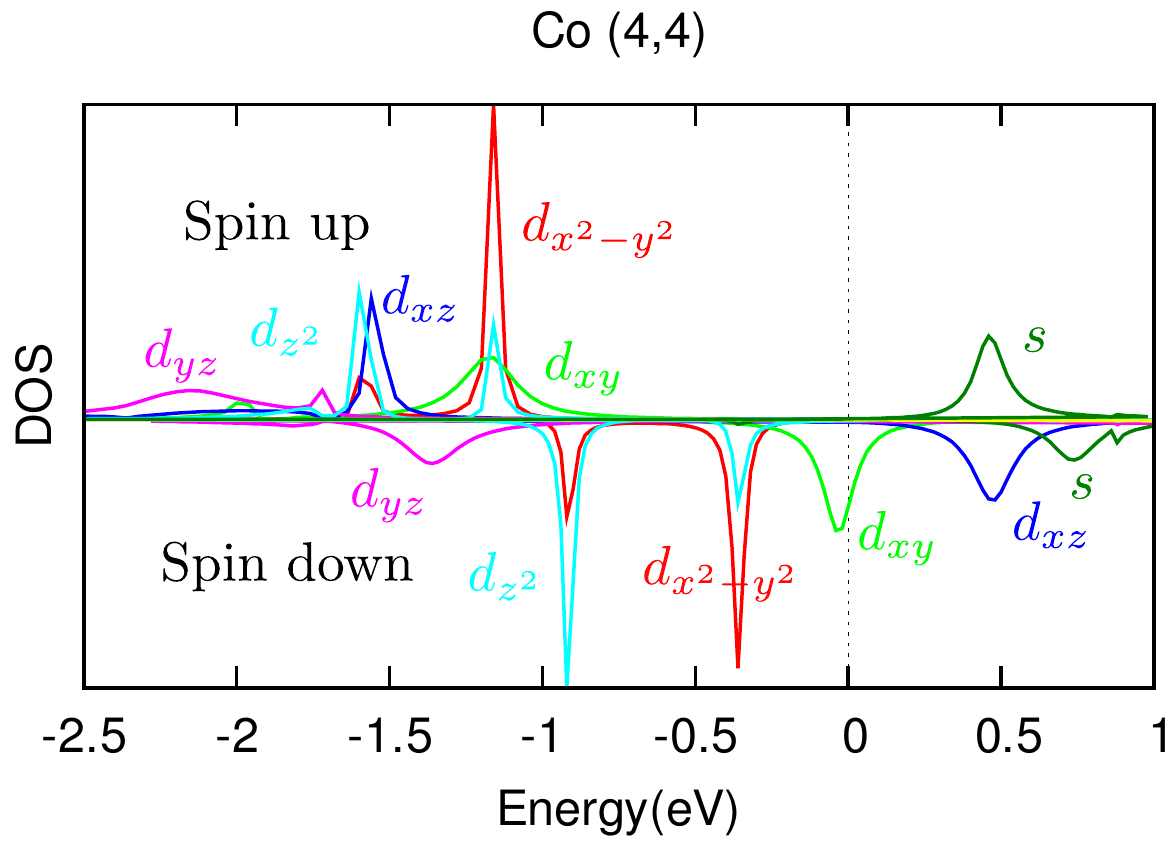}
\includegraphics[width=0.45\textwidth]{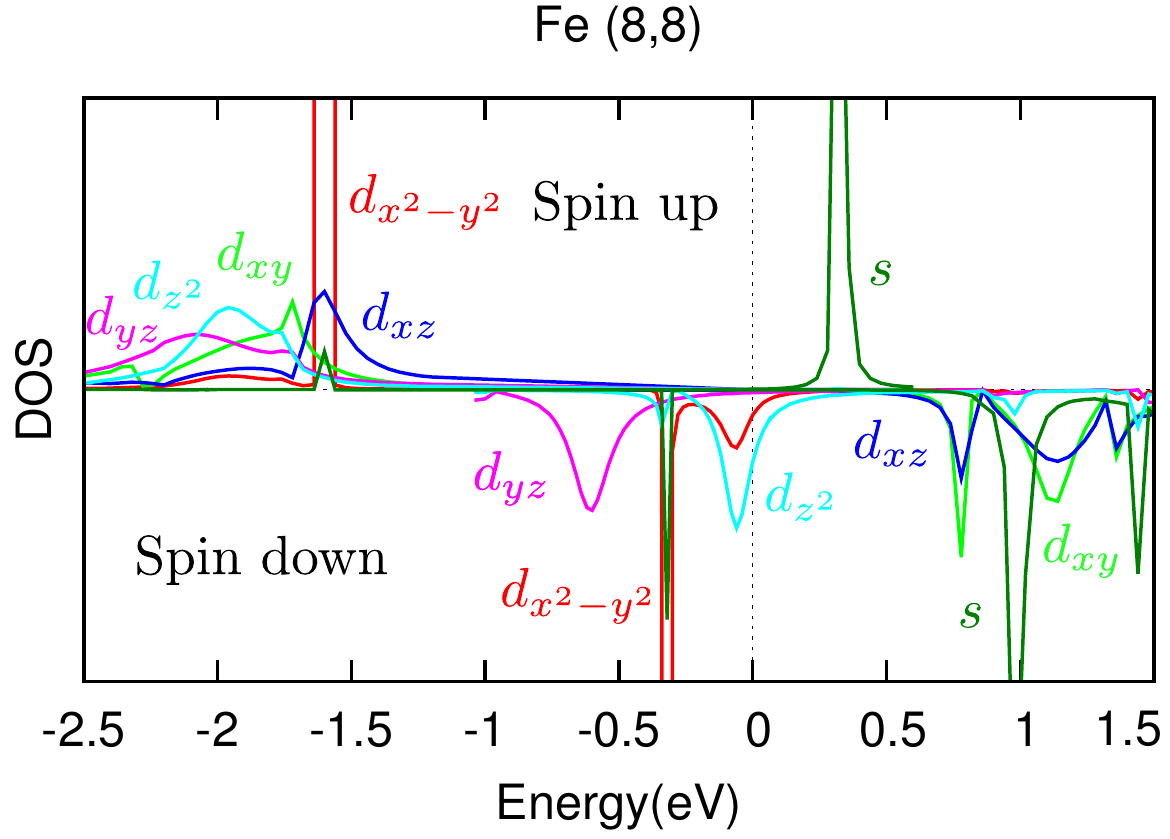}
\caption{Projected density of states of impurity $3d$ and $4s$ orbitals. Above: Co on (4,4) SWNT; below: Fe on (8,8) SWNT.}\label{fig3}
\end{figure}

We found that Co behaves as a $S=\frac{1}{2}$ impurity on both (4,4)
and (8,8) SWNTs, its $d_{xz}$ orbital driving the spin polarization. The 
Co atom switches from the  $3d^74s^2$ configuration of the isolated atom 
to a slightly surprising low-spin $3d^94s^0$ one when adsorbed on the nanotube. 
Fe behaves as a $S=1$ impurity on the (8,8) tube, similarly switching from 
the high-spin $3d^64s^2$ of the isolated atom to a low-spin $3d^84s^0$ in
the adsorbed state. Here the pair of orbitals $d_{xz}$ and $d_{xy}$ is magnetically 
polarized (see fig. \ref{fig3} and tab. \ref{tab2}), a result in
good agreement with previous calculations \cite{durgun, yagi}. Orbitals
$d_{xy}$ for Co and $d_{z^2}$ for Fe are partly empty, and fall near
the Fermi energy in straight DFT: but they promptly move below
$E_F$ when even a small $U$ is switched on. We conclude that these orbitals are not 
going to be involved in Kondo behaviour and can be neglected to a first approximation 
in order to keep the many-body model simple.
The behavior of Fe/(4,4) is complicated. The $s$ orbital is partly filled, 
and $d_{z^2}$ is magnetically polarized besides the ($d_{xz}$,$d_{xy}$) pair, 
so here Fe should behave as a $S=\frac{3}{2}$ impurity.

\begin{figure}\centering
\includegraphics[width=0.33\textwidth, angle=270]{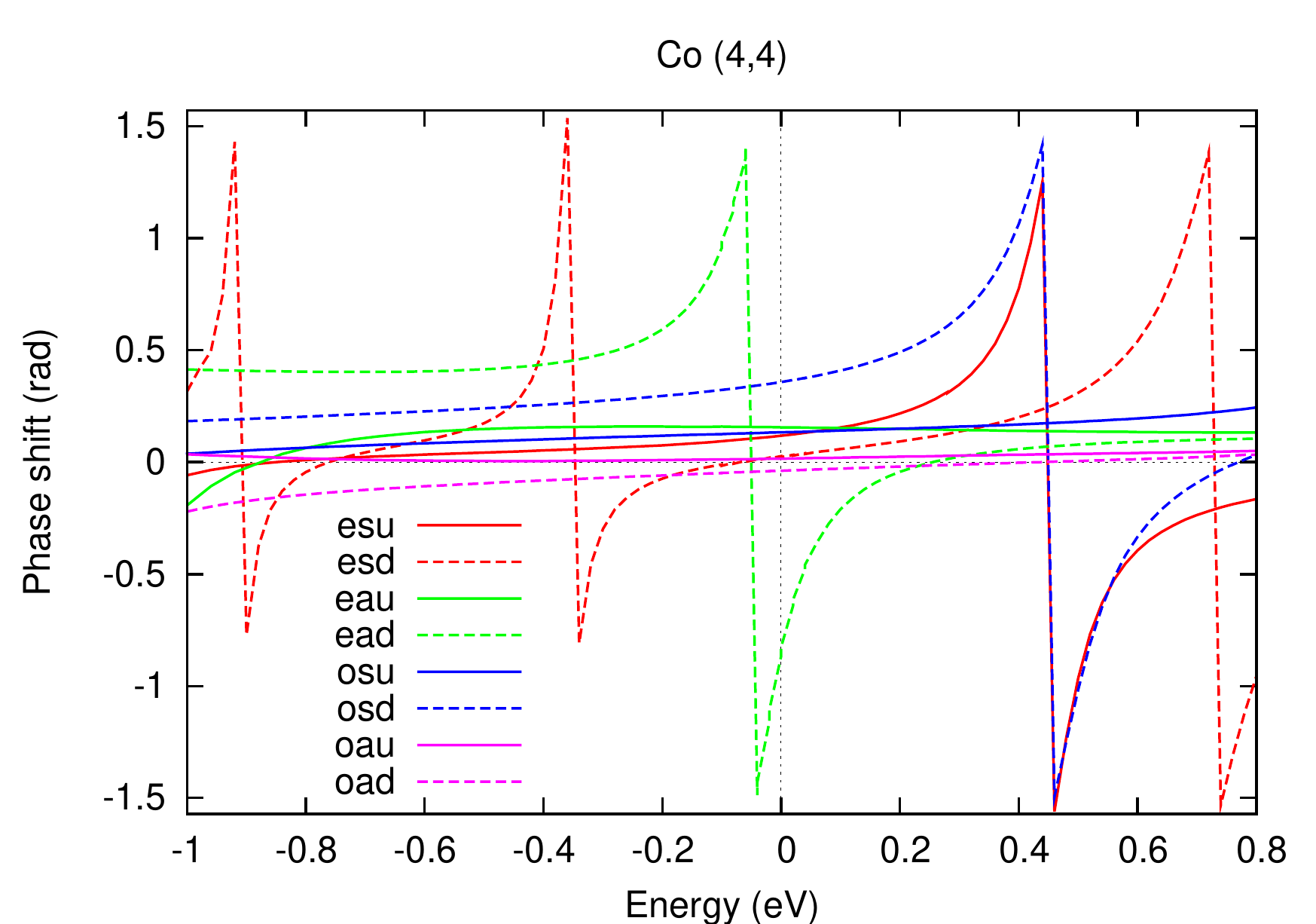}
\includegraphics[width=0.33\textwidth, angle=270]{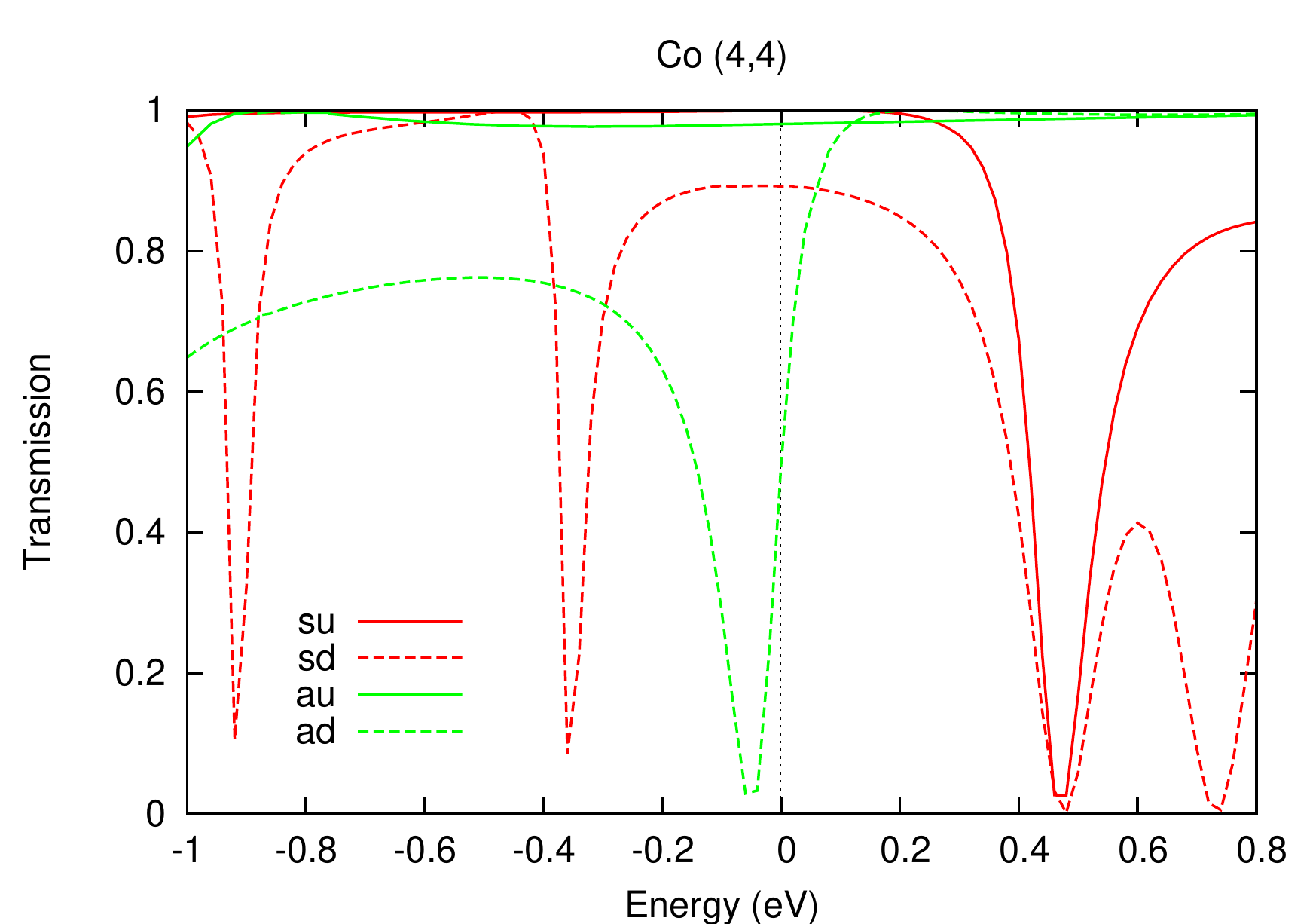}
\caption{Above: conductance as a function of energy of conduction electrons for Co on (4,4) SWNT, for each symmetry $s-a$ and spin direction $u-d$ (up-down); below: phase shift of conduction electrons for different symmetries $s-a$, $e-o$ and spin directions $u-d$.}\label{fig4}
\end{figure}

As in previous work \cite{naturemat} we implemented DFT computation of
the (spin-polarized) mean-field ballistc conductance and, more importantly, 
of the impurity-related spin- and channel-selected phase shifts suffered by the SWNT conduction 
electrons as a function of energy. An example is shown in fig. \ref{fig4} for Co on the (4,4) SWNT.

\section{Generalized Anderson model}\label{sec4}
The Kondo model is usually understood by means of a many-body Anderson Hamiltonian \cite{anderson61}. 
In our case we need to extend it in principle to the four SWNT conduction bands, each hybridized 
with some impurity orbital among the $3d$ and $4s$, of same symmetry. These impurity orbitals 
in turn are  mutually coupled by an intra-atomic ferromagnetic Hund exchange term,
\begin{equation}\label{hamtot}
H=\sum_{i=es, ea, os, oa} H^{And}_i+H_{Hund}
\end{equation}
\begin{equation}
 H_{Hund}=J\sum_{i<j, j=1,4}  \vec{{\sigma}}_i\cdot\vec{{\sigma}}_j
\end{equation}
\begin{equation}
\vec{\sigma}_i=\frac{1}{2}\sum_{\alpha\beta}d_{i\alpha}^\dagger \vec{s}_{\alpha\beta} d_{i\beta}
\end{equation}
\begin{equation}\label{h_and}
H^{And}_i=\sum_k\epsilon_{ik}c^\dagger_{ik}c_{ik}-t_ic^\dagger_ic_i+ V_i (c_i^\dagger d_i+d_ic^\dagger _i)+\epsilon_i d_i^\dagger d_i+U_in_i^\uparrow n_i^\downarrow
\end{equation}
where $d^\dagger_i$ creates an electron on the impurity orbital with symmetry $i$, $c^\dagger_{ik}$ 
creates an electron in a $k$ conduction state with symmetry $i$, $\vec{s}$ are the Pauli matrices, 
$t_i$ is a potential scattering term due to to the charge density of the impurity, 
$V_i$ is the coupling of the impurity orbital with the conduction electron states, 
$\epsilon_i$ is the bare energy of the impurity orbital, $U_i$ is the Hubbard 
repulsion on the orbital, $J<0$ is a global Hund exchange parameter (favouring high spin for the
isolated impurity), and the single particle energies of conduction electrons $\epsilon_{ik}$ 
are such as to give a constant density of states, with exactly the same value as that of the 
clean SWNT (per spin direction) as computed by DFT:
\begin{equation}
 \sum_k\delta(\epsilon-\epsilon_{ik})=\rho\simeq\frac{1}{12\mbox{eV}},\hspace{10pt}i=es,eo, as, ao
\end{equation}

In practice, only conduction bands coupled to a magnetic orbital are retained in our NRG procedure 
(see section \ref{sec6}). This leaves us with a single band coupled to a single impurity level in 
the case of Co (once orbital $d_{xy}$ is ignored), and with two bands, each coupled to one impurity level, 
in the case of Fe/(8,8)  (once orbital $d_{z^2}$ is ignored).
The case of Fe/(4,4) is more involved and we will presently not deal with it.

\section{Joining up DFT and many body}\label{sec5}
Hamiltonian eq. \ref{hamtot} is easily solved in the (unrestricted) Hartree-Fock approximation 
\cite{anderson61}, breaking spin rotational symmetry. This leads to a phase shift in conduction 
electrons of symmetry $i$ ($i=es, eo, as, ao$) at the Fermi energy
\begin{equation}
 \delta_i^{\sigma}=\phi_i+\arctan\frac{\Gamma_i}{\epsilon_i^{\sigma }}
\end{equation}
where $\phi_i=\arctan\pi\rho t_i\simeq 0$ is the phase shift caused by the impurity charge 
scattering. This is numerically found to be negligible, so we shall ignore it from now on. 
The peak of the impurity DOS is found to be at
\begin{equation}
\epsilon_i^{\sigma}=\epsilon_i+U_i\langle n_i^{-\sigma}\rangle-\sigma\frac{J}{2}(m_{tot}-m_i)
\end{equation}
where
\begin{equation}
 \langle n_i^{\sigma}\rangle=\frac{1}{\pi}\arctan\frac{\Gamma_i^\sigma}{\epsilon_i^{\sigma}}
\end{equation}
is the average occupation of up/down orbital 
\begin{equation}
 m_i=\langle n_i^\uparrow\rangle-\langle n_i^\downarrow\rangle
\end{equation}
is the magnetization of each orbital and
\begin{equation}
 m_{tot}=\sum_{i=1}^{n}m_i
\end{equation}
is the total magnetization of the atom.
As in \cite{naturemat}, we choose to reproduce the same phase shifts at the Fermi 
energy for each symmetry, and the same peaks in the density of states of the impurity orbitals 
as those computed by DFT.
This allows to uniquely fix $\epsilon_i$, $U_i$ and $\Gamma_i$ as long as just one magnetic 
orbital is considered in eq. \ref{hamtot} -- that is the case of Co ($i=os$). When 
more than one orbital is involved, such as in Fe, (or in Co if orbital $d_{xy}$ were to be 
taken into account) we need to fix $J$ as well. We can extract J from the DFT calculated  
exchange splitting of filled orbitals, according to 
\begin{equation} 
\epsilon_f^{\uparrow}-\epsilon_f^{\downarrow}=\frac{J}{2}m_{tot}
\end{equation}
Since different $d$ orbitals have slightly different splittings, we just took an average 
value as deduced form different orbitals. This yields $J\sim 1$ eV in Co and $J\sim 1.2$ eV in Fe.

\section{Results of NRG calculations}\label{sec6}

We solved the Anderson Hamiltonian by means of NRG \cite{wilson, bulla08}, which allows 
to compute all the needed static and dynamic quantities we need in an almost exact, albeit
numerical, way. We extracted the conduction electrons phase shifts from the single particle 
energies at the zero energy fixed point, and the Kondo temperature from the impurity Green 
function at imaginary frequency:
\begin{equation}
G(i\epsilon)=\frac{1}{i\epsilon-\epsilon_i-\Sigma(i\epsilon)+i\Gamma_i}=\frac{1}{Z_{part}}\sum_{n}\frac{|\langle GS|d|n\rangle|^2}{i\epsilon-\epsilon_n}
\end{equation}
($Z_{part}$ is the partition function and $GS$ the ground state). The Kondo temperature is given by
\begin{equation}
 T_K=\frac{\pi wZ\Gamma}{4k_b}
\end{equation}
where $w=0.4128$ is the Wilson coefficient and $Z$ is the quasiparticle residue
\begin{equation}
 Z^{-1}=1-\frac{\partial\Sigma(i\epsilon)}{\partial(i\epsilon)}
\end{equation}
Alternatively, an approximate formula \cite{hewson}, valid for one impurity coupled to one channel,
\begin{equation}\label{tkform}
 T_K\sim0.4107\sqrt{\frac{U_i\Gamma_i}{2}}e^{\pi\epsilon_i(\epsilon_i+U_i)/2\Gamma_iU_i}
\end{equation}
could be used, 
with similar results. 

The zero-bias conductance is given, in terms of the final phase shifts, by
\begin{equation}
g\equiv \frac{G}{G_0}=\cos^2(\delta_{es}-\delta_{os})+\cos^2(\delta_{ea}-\delta_{oa})\equiv g_s+g_a
\end{equation}
where $G_0\equiv\frac{2e^2}{h}$ is the quantum of conductance. Note that in the clean tube $G=2G_0$. 
Phase shifts are only computed for Kondo channels, and are found to be always $\simeq \pi/2$. 
For the non-Kondo channels, they can be directly extracted from DFT. Since in DFT they are $\simeq 0$, 
they can be safely neglected. Summing up, both Co/(4,4) and Co/(8,8) should exhibit a 
(zero temperature and zero bias) conductance $G\sim G_0$,  whereas Fe/(8,8) should have $G\sim 0$. 
These results remain valid so long as either temperature and/or bias remain well below
$T_K$. However, it turns out that Kondo temperatures $T_k$ are quite low (see tab. \ref{tab3}), 
which might make this effect hard to observe in a real experiment.
Interestingly, a much higher Kondo temperature of about 15 K has been quoted for Co/graphene\cite{lichtenstein}.
While the reasons for this difference between graphene and nanotubes are presently 
being investigated, it should be noted that several factors differ, including symmetry,
and heavy doping in real, deposited graphene.

\begin{figure}\centering
\includegraphics[scale=0.4, angle=270]{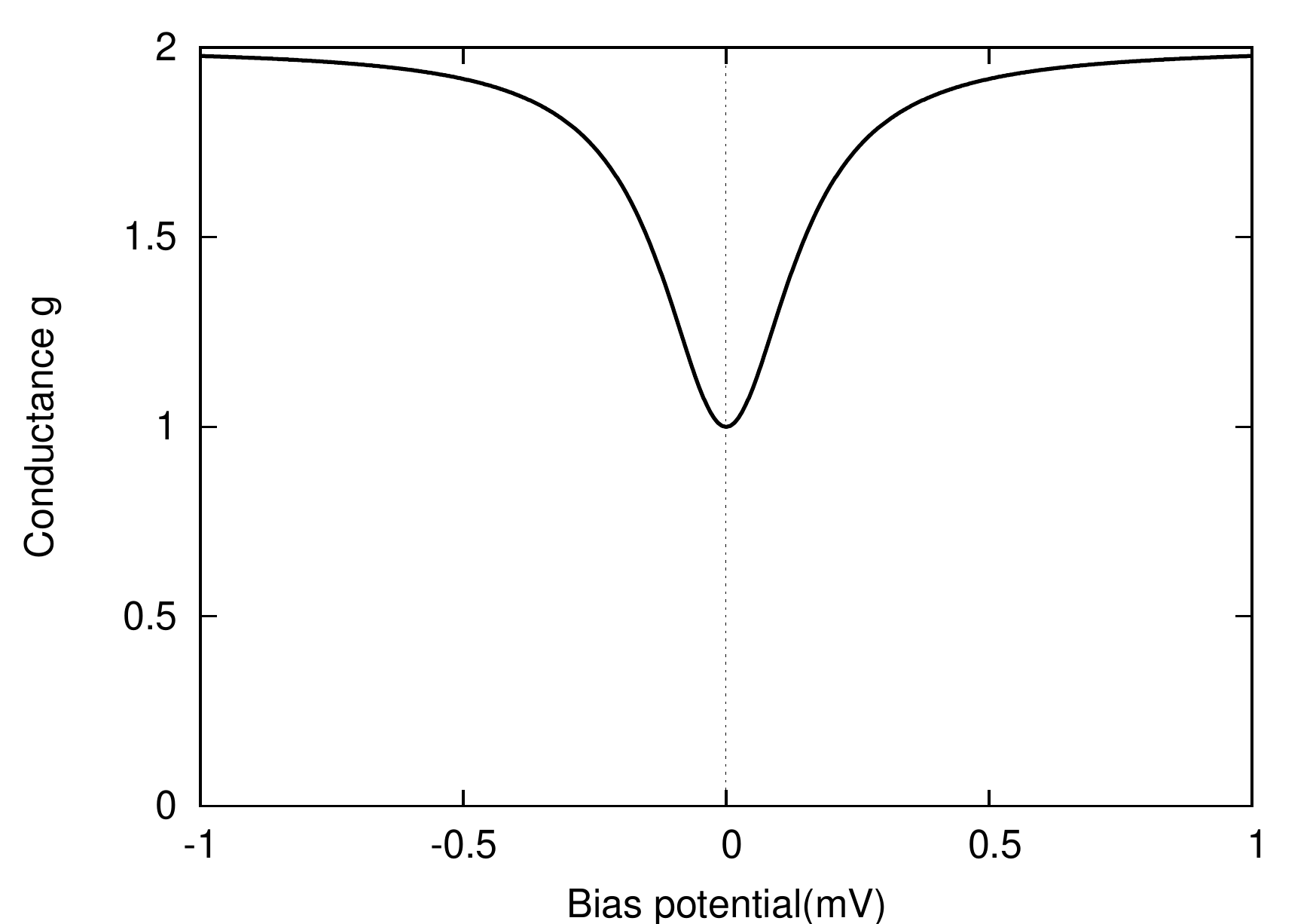}
\includegraphics[scale=0.4, angle=270]{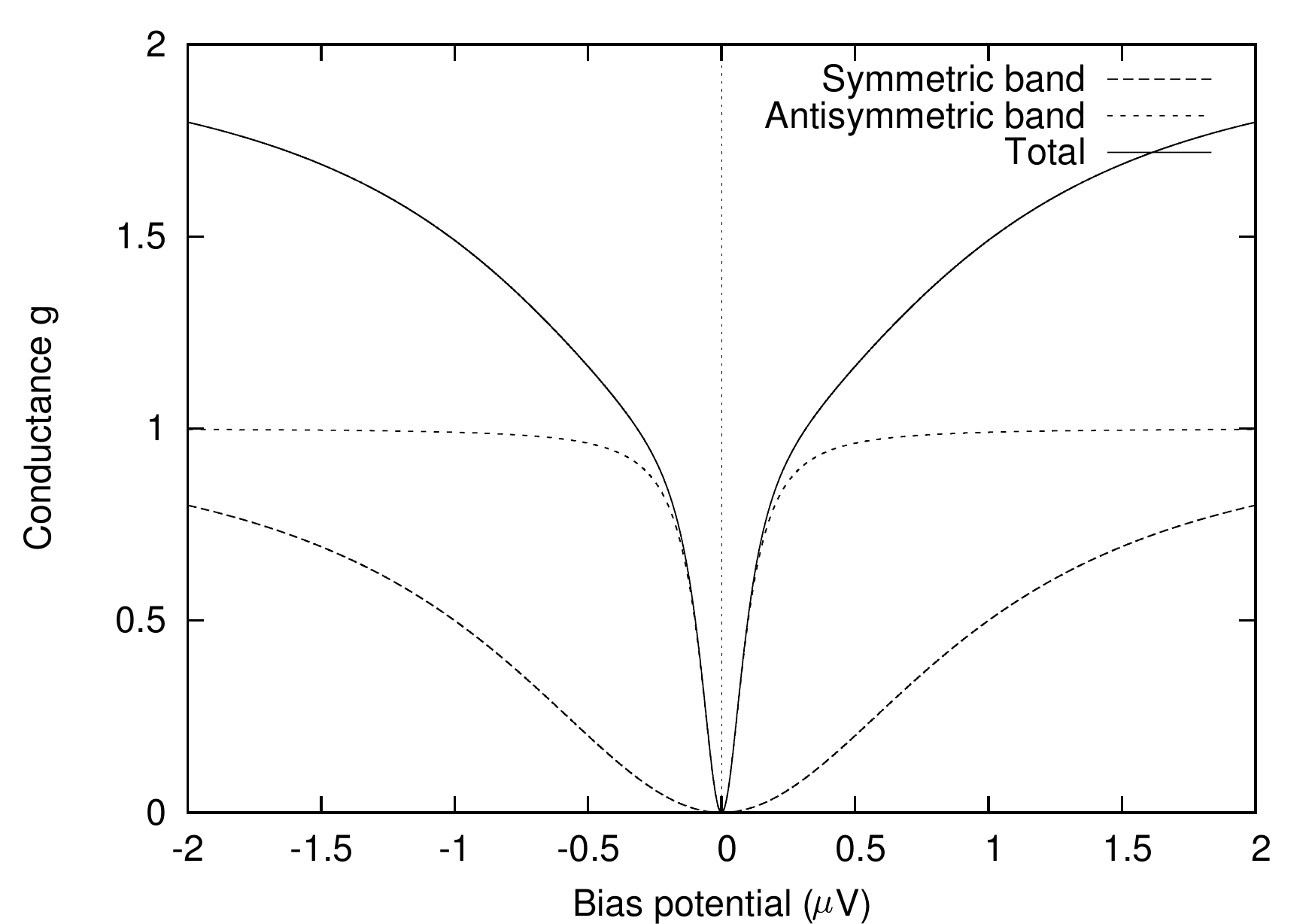}
\caption{Predicted zero-bias anomaly for Co impurity on both (4,4) and (8,8) SWNTs (above) and for Fe on (8,8) SWNT (below) ($g\equiv G/G_0$, $G_0=2e^2/h$). } \label{fig5}
\end{figure}

Finally, we can qualitatively address the predicted bias-dependent lineshape of the Kondo 
conductance anomaly. Through the Keldysh technique for non-equilibrium Green-functions it 
is possible to compute the finite-bias conductance \cite{meir}, once the impurity Green 
function $G_i(\epsilon)$ is calculated from NRG \cite{costi94}:
\begin{equation}
 g_{s,a}=1-\Gamma\Im G_i(\epsilon)
\end{equation}
For simplicity, we have taken
\begin{equation}\label{kondopeak}
 G_i(\epsilon)=\frac{\Gamma_k/\Gamma}{\epsilon+i\Gamma_k}
\end{equation}
where
\begin{equation}
k_bT_K=\frac{w\pi}{4}\Gamma_K=0.342\Gamma_k
\end{equation}
This gives rise to a Fano lineshape \cite{fano}
\begin{equation}
g_{s,a}=\frac{(q+v)^2}{(q^2+1)(v^2+1)},\hspace{20pt}v\equiv\epsilon/\Gamma_k 
\end{equation}
with $q=0$, so for each band ($s-a$) the lineshape is predicted to be a symmetric 
antilorentzian, with a width proportional to the Kondo temperature. Small sources 
of asymmetry will arise from a) the potential scattering $t_i$ which we ignored in 
Eq. \ref{h_and};b) from the interference with orbitals belonging to the same band 
$a$ or $s$, but with different symmetry $e/o$;  and c) from particle-hole asymmetries 
in eq. \ref{kondopeak}. However, we estimate that the asymmetry parameter $q$ should 
generally remain below 0.1. In Co/(4,4) and Co/(8,8), only $g_s$ contributes to the 
lineshape, $g_a$ being almost one -- and moreover independent from energy on the 
Kondo energy scale.  In Fe/(8,8), both $g_s$ and $g_a$ have an antilorentzian shape,
although with very different widths. The total lineshape is just their sum (see fig. \ref{fig5}).

\section{Conclusions}\label{sec7}

We implemented our recently devised DFT+NRG scheme \cite{naturemat} to calculate the Kondo effect caused by
Co and Fe adsorbed impurities on the conductance of (4,4) and (8,8) nanotubes.  On the
methodological side, the present calculation represents a good pedagogical illustration
of our technique. For the systems chosen, the predicted anomalies are symmetric 
antilorentzian dips, reducing total zero bias conductance to zero for Fe, and 
by a factor 1/2 for Co. While there are no data to compare with, this prediction should in
principle be amenable to experimental check. However, we note that our calculated Kondo
temperatures are very small, which might constitute and experimental challenge.

This work was sponsored under PRIN/COFIN, and also under FANAS/AFRI.
We are indebted to A. Ferretti, P. Lucignano, R. Mazzarello, and L. De Leo for early collaboration
and discussions.


\begin{table}\centering
\begin{tabular}{c|c|c}
\hline& s&a\\
\hline e&$d_{z^2}$,$d_{x^2-y^2}$,$s$&$d_{xy}$\\
\hline o&$d_{xz}$&$d_{yz}$ 
\end{tabular}
\caption{Symmetries of $d$ and $s$ orbitals with respect to the $xy$ plane (even $e$ - odd $o$) and the $xz$ plane (symmetric $s$ - antisymmetric $a$).}\label{tab1}
\end{table}

\begin{table}\centering
\begin{tabular}{c|c|c|c}
\hline Impurity& Magnetic Orbital&Symmetry& Spin\\
\hline Co&$d_{xz}$&$os$&$\frac{1}{2}$\\
 &($d_{xy}$)&$ea$&0\\
\hline Fe&$d_{xz}$&$os$&$\frac{1}{2}$\\
&$d_{xy}$&$ea$&$\frac{1}{2}$\\ 
&($d_{z^2}$,$s$)&$es$&0\\ 
\end{tabular}
\caption{Magnetic orbitals as found from DFT calculations, their symmetry, and spin they carry ($S=1/2$ for each magnetic orbital). Orbitals in parentheses are not Kondo orbitals, so do not contribute to the total spin of the impurity and are ignored in the many-body model, but still participates in transport.}\label{tab2}
\end{table}

\begin{table}\centering
\begin{tabular}{c|c|c|c|c|c|c}
\hline Imp.&SWNT &Orb.&$\epsilon_d$(eV)&U(eV)&$\Gamma$(eV)&$T_K$(K)\\
\hline Co&(4,4)&$d_{xz}$&-1.62&2.17&0.082&$\sim 1$\\
\cline{2-7}&(8,8)&$d_{xz}$&-1.83&2.11&0.054&$\sim 1$\\
\hline Fe&(8,8)&$d_{xz}$&-1.24&2.01&0.060&$\sim 10^{-4}$\\
\cline{3-7}&&$d_{xy}$&-1.38&2.13&0.043&$\sim 10^{-3}$\\ 
\end{tabular}
\caption{Recapitulative table of important quantities of our Anderson models. Kondo temperature is so low in Fe due to the reduced broadening $\Gamma$ and to the Hund coupling $J$ that couples the two impurity orbitals.}\label{tab3}
\end{table}


\end{document}